\documentstyle[11pt,newpasp,twoside,epsf]{article}
\def\edcomment#1{\iffalse\marginpar{\raggedright\sl#1\/}\else\relax\fi}
\reversemarginpar

\begin{document}
\title{The L$_x$ - $\sigma$ Relation For Poor Groups of Galaxies}
 \author{Marc E. Zimer}
\affil{Max Planck Institute for Astronomy, Heidelberg}
\author{John S. Mulchaey}
\affil{Carnegie Observatories, Pasadena}
\author{Ann I. Zabludoff}
\affil{Steward Observatory, University of Arizona, Tucson}

\begin{abstract}

We use a sample of 20 poor groups of galaxies to study the low L$_x$
tail of the L$_x$ - $\sigma$ relationship. We have obtained redshifts
for fainter members and deep X-ray imaging of these groups. We find
that the L$_x$ - $\sigma$ relationship derived for the rich clusters
in the sample of Mushotzky \& Scharf (1997) is consistent with the
data for our 20 poor groups. However, it is not possible to reach
strong conclusions about differences in the L$_x$ - $\sigma$
relationship for groups and clusters in general at the present time
because of significant differences between various cluster samples.

\end{abstract}

\section{Introduction}
Poor groups of galaxies, such as the Local Group are the most common
environment for galaxies. ROSAT observations indicate that
approximately half of all nearby groups of galaxies contain a diffuse,
X-ray emitting `intragroup medium' analogous to the intracluster
medium in rich clusters (Mulchaey 2000). Recently it has been realized
that the intragroup medium may hold important clues into the formation
and evolution of groups (see for example, Dav\'e et al, these
proceedings).  Numerical simulations indicate that in the absence of
non-gravitational heating, the relationships between X-ray luminosity
(L$_x$), X-ray temperature (T) and optical velocity dispersion
($\sigma$) are expected to be similar for groups and clusters. The
true nature of the scaling relationships for groups and clusters has
proved to be a very controversial topic. For example, different
authors have reached vastly different conclusions about the
relationship between the optical velocity dispersion $\sigma$ and the
X-ray luminosity L$_x$. Previous work has found either a similar
(Zabludoff $\&$ Mulchaey 1998) or a more shallow (Helsdon $\&$ Ponman
2000; Mahdavi et al. 2000) relation for groups than for clusters.
Several explanations for a shallower group slope have been proposed
including the possibility that individual galaxy halos contribute
significantly to the X-ray luminosity of groups (Mahdavi et al. 2000).

Most studies of the L$_x$-$\sigma$ relationship for groups have
been based on velocity dispersions determined from only 3 or 4 of the
brightest group members. Simulations indicate that when small numbers
of galaxies are used, the velocity dispersion does not trace the group
mass (Dav\'e et al., these proceedings).  We have embarked on a study
to measure velocities for more group members and therefore obtain
accurate dispersions for a large number of groups.

\section{The L$_x$ - $\sigma$ Plot}
Our sample contains 20 groups of galaxies observed with fiber
spectrographs at the 2.5m telescope at Las Campanas Observatory and
the 3.5m WIYN telescope (Zabludoff \& Mulchaey 1998; Zimer et
al. 2001).  The X-ray properties of these groups have been derived
from ROSAT PSPC data (Mulchaey et al.  2001).  Fourteen of these
groups are detected in the ROSAT images. From the fiber spectroscopy,
we measure velocities for typically $\sim$ 15--50 group members.
These numbers are high enough to derive reliable velocity dispersions
(Dav\'e et al., this proceedings). Figure 1 shows the resulting
L$_x$-$\sigma$ relationship we derive (data points), with
various other fits to the L$_x$-$\sigma$ relationship found in
the literature.  A summary of the various published L$_x$-$\sigma$
relationships is also given in Table 1.  The $upper$ $left$ panel in
Figure 1 shows the best fit to our cluster and group data using a
parametric bootstrap technique (Lubin $\&$ Bahcall 1993):

\begin{equation}
log\mbox{ } L_x = (4.39\pm0.27)\mbox{ } log\mbox{ } \sigma + (31.29\pm0.80)
\end{equation}

while for the cluster data alone (Mushotzky \& Scharf 1997),
we derive: 
\begin{equation}
log\mbox{ } L_x = (3.67\pm0.51)\mbox{ } log\mbox{ } \sigma + (33.46\pm1.54).
\end{equation}
Within the errors, these fits are in agreement with each other and
with the earlier results of Mulchaey $\&$ Zabludoff (1998) (see Table
1).  Given the small number of groups in our sample, we have not
attempted to fit the group data alone. However, a visual inspection of
Figure 1 suggests the group + cluster fit adequately describes the
groups.
In the $upper$ $right$ panel we compare our data with fits in Xue $\&$
Wu (2000).  Their group + cluster fit is consistent with our result
within the errors.  The $lower$ $left$ panel shows the best fit slopes
for groups only from Mahdavi et al.  (2000) and Helsdon \& Ponman
(2000).  It is immediately obvious that the Mahdavi et al.  (2000) fit
is inconsistent with our data and cannot be correct for groups in
general.  The Helsdon $\&$ Ponman (2000) group fit is consistent with
our group + cluster fit.  However, Helsdon $\&$ Ponman (2000) claim to
find a significant difference between the L$_x$-$\sigma$
relation for groups and the relation for clusters. Despite finding
similar slopes for groups, we reach a different conclusion than
Helsdon $\&$ Ponman (2000) because there are significant differences
in the cluster samples used in each survey.  Helsdon $\&$ Ponman
(2000) use the White et al. (1997) cluster sample (see the $lower$
$right$ panel in Figure 1), which has a considerably steeper L$_x$-$\sigma$ 
relation than the Mushotzky $\&$ Scharf (1997) clusters ample we use (White 
et al. 1997 : L$_x$ $\propto$ $\sigma$$^{6.38}$, Mushotzky $\&$ Scharf 1997 
: L$_x$ $\propto$ $\sigma$$^{3.67}$). The reason for the differences in 
these cluster samples is not apparent. It is clear, however, that these 
differences must be understood before any strong conclusions can be made 
regarding differences in the L$_x$-$\sigma$ relationship of groups and
clusters.

\begin{figure}
\small
\plotone{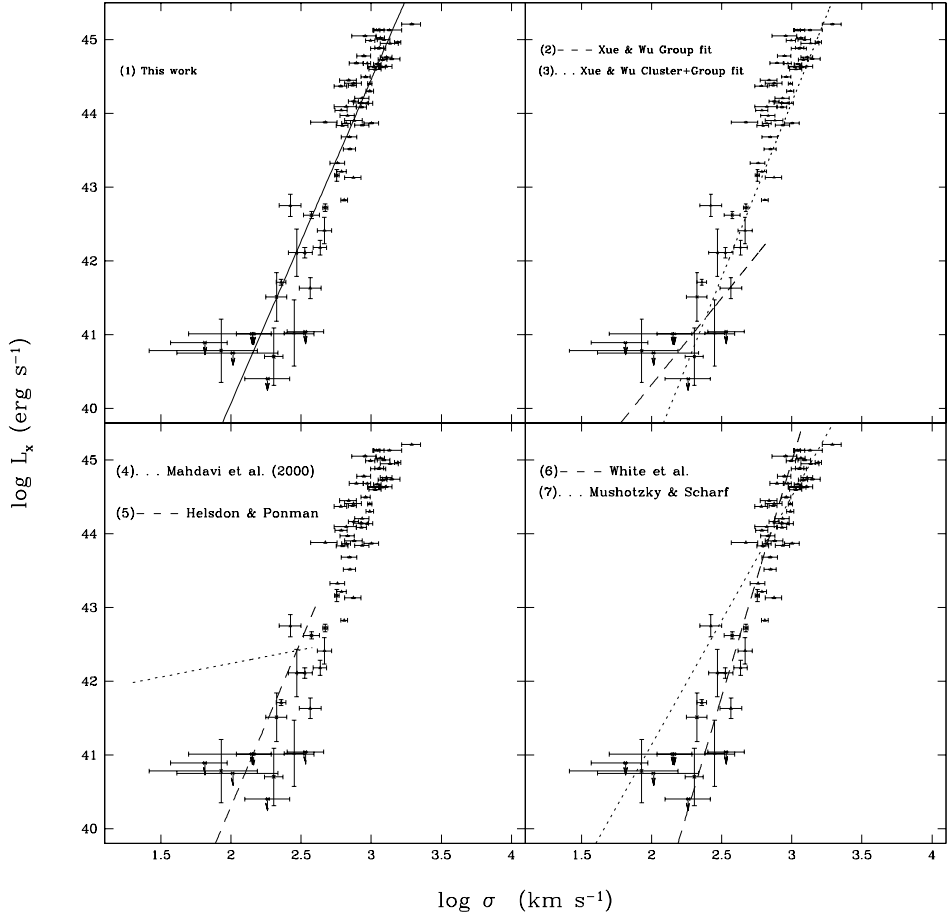}
\caption{Logarithm of velocity dispersion vs. logarithm of X-ray luminosity
for the X-ray detected groups and clusters. The non-detected groups are 
plotted using the upper limits on L$_x$ (arrows) 
}
\end{figure}
\begin{table}
\small
\caption{The L$_x$ - $\sigma$ Relation fits}
\begin{tabular}{|c||c|c|c|}
\hline
 & \textbf{Helsdon $\&$ Ponman 2000} &  \textbf{Mahdavi et al. 2000} &  \textbf{Xue $\&$ Wu 2000} \\ \hline \hline
groups only& 4.7$\pm$0.9 (5)& 0.37$\pm$0.3 (4)& 2.35$\pm$0.21 (2)\\ \hline 
clusters only& 6.38$\pm$0.46 (6)& 3.90$\pm$0.10 & 5.30$\pm$0.21 \\ \hline
clusters + groups & / & / & 4.75$\pm$0.18 (3) \\ \hline \hline 
\hline
&  \textbf{Zabludoff $\&$ Mulchaey 1998} &  \textbf{This work} & \\ \hline \hline
groups only& / & / & \\ \hline
clusters only& / & 3.67$\pm$0.51 (7) & \\ \hline
clusters + groups & 4.29$\pm$0.37 & 4.39$\pm$0.27 (1) & \\ \hline
\end{tabular}
\end{table}
\section{Conclusion}
There is still considerable debate about the nature of the L$_x$ -
$\sigma$ relationship for groups and clusters.  Some of the
discrepancies in the literature may be due to poorly determined
$\sigma$'s and/or L$_x$'s. In many studies, the $\sigma$'s are quite
uncertain because they were derived from only the three or four
brightest group members. The L$_x$ measurements may also be uncertain
because of the low signal-to-noise of most ROSAT images and
contamination from other sources.

From our fiber spectroscopy program we have found: 
\begin{itemize}
\item low-mass groups have many members if the luminosity 
function is sampled down to sufficiently low luminosities. 
\item robust velocity dispersions for 
groups require at least $\sim$ 10--15 velocity measurements.
\item strong conclusions about differences in the L$_x$ - $\sigma$
relationship for groups and clusters cannot be reached at the present
time because of differences in the various cluster samples.
\item the uncertainties in the various scaling relationships suggest that it
may be somewhat premature at this time to draw strong conclusions about the 
need for additional heating mechanisms in groups.

\end{itemize}

\end{document}